# Microscopic driving theory with non-hypothetical congested steady state: model and empirical verification

Jun-fang Tian∗[1], Martin Treiber[2], Bin Jia[3], Wen-yi Zhang[3]

[1]*Institute of Systems Engineering, College of Management and Economics, Tianjin University, No. 92 Weijin Road, Nankai District, Tianjin 300072, China*

[2]*Technische Universität Dresden, Institute for Transport & Economics, Würzburger Str. 35, D-01062 Dresden, Germany*

[3]*MOE Key Laboratory for Urban Transportation Complex Systems Theory and Technology, Beijing Jiaotong University, Beijing, 100044, China*

The essential distinction between the fundamental diagram approach and three-phase theory is the existence of the unique space-gap-speed relationship. In order to verify this relationship, empirical data are analyzed with the following findings: (1) linear relationship between the actual space gap and speed can be identified when the speed difference between vehicles approximates zero; (2) vehicles accelerate or decelerate around the desired space gap most of the time. To explain these phenomena, we propose that, in homogeneous congested traffic flow, the space gap between two vehicles will oscillate around the desired space gap in the noiseless limit. This assumption is formulated in terms of a cellular automaton. Simulations under periodic and open boundary conditions reproduce the empirical findings of three-phase theory. Finally, the model is calibrated and validated. All verification results are acceptable and better than that of previous studies.

## 1. Introduction

With the rapid development of urbanization, traffic congestion becomes one of the most serious problems that undermine the operational efficiency of modern cities. In order to understand the mechanism of traffic congestion, many models and analysis have been carried out to explain the empirical findings (see the reviews: Haight, 1963; Whitham, 1974; Leutzbach, 1987; Chowdhury, 2000; Helbing, 2001; Nagatani, 2002; Jia et al. 2007; Kerner, 2004, 2009; Treiber and Kesting, 2013). Generally speaking, these models can be classified into the fundamental diagram approach and models consistent with three-phase theory.

Originated from Greenshields (1935), the fundamental diagram permeates all levels of traffic flow models and is one of the basic research technics of empirical data. The fundamental diagram is the idealized form of the flow-density curve in traffic flow, which goes through the origin and has at least one maximum. It describes the theoretical relationship between density and flow in the stationary homogeneous traffic, i.e., the steady state of identical driver-vehicle units (Treiber and Kesting, 2013). In the last century, almost all traffic flow models belong to the fundamental diagram approach. In microscopic models, the fundamental diagram is linked to the steady states of car-following (CF) or cellular automaton (CA) models. For example, in the Optimal Velocity Model (OV model) by Bando et al. (1995), the fundamental diagram corresponds to the optimal velocity function itself. In the Nagel-Schreckenberg cellular automaton model (NaSch model), it could be derived in terms of the steady state in the deterministic limit (Nagel and Schreckenberg, 1992). In macroscopic or mesoscopic models, it has been directly applied (e.g. the LWR theory (Lighthill and Whitham, 1955; Richards, 1956)) or incorporated into the momentum equation (e.g. the PW theory (Payne, 1979)).

---

*Corresponding author.
E-mail address:jftian@tju.edu.cn

The majority of models within the fundamental diagram approach belongs to the two-phase models (Lighthill and Whitham, 1955; Richards, 1956; Herman et al., 1959; Payne 1979; Gipps, 1981; Nagel and Schreckenberg, 1992; Daganzo, 1994; Bando et al., 1995; Krauss et al., 1997; Treiber et al., 2000; Aw and Rascle, 2000; Newell, 2002; Tang, 2005)), which refers to the free flow phase (F) and the jammed phase (J). The phase transitions involved are the transition from free flow to jams (F→J transition) and the transition from jam to free flow (J→F transition). The fundamental diagram approach explains the jam formation mainly by excess demand, i.e., the traffic inflow exceeds the static capacity defined by the maximum of the fundamental diagram. Additionally, instabilities of traffic flow, which are caused by finite speed adaption times (due to finite accelerations) or reactiontimes, can lead to jam formation even before static capacity is reached. For the detailed discussion of stability, one can refer to Treiber and Kesting (2013), Kesting and Treiber (2008).

Based on the long-term empirical analysis, Kerner (2004, 2009) argues that two-phase models could not reproduce the empirical features of traffic breakdown as well as the further development of the related congested region properly. Therefore, the three-phase theory is introduced distinguishing (1) free traffic flow, (2) synchronized flow, and (3) wide moving jams. The fundamental hypothesis of the three-phase theory is that the hypothetical steady states of the synchronized flow cover a two-dimensional region in the flow-density plane. In other words, there is no fundamental diagram of traffic flow. Over the time, many models within the framework of three-phase theory are proposed (Kerner et al., 2002, 2011; Kerner, 2012; Kerner and Klenov 2002, 2003, 2006; Lee et al., 2004; Jiang and Wu, 2003, 2005; Tian et al., 2009; Gao et al., 2007, 2009; Davis, 2004). In order to improve readability, we have made a brief introduction of three-phase theory in the appendix.

It should be noted that there are models within the fundamental diagram approach that could reproduce the three-phase theory, such as the Brake Light cellular automaton Model (BLM (Knospe, 2000)), the Speed Adaption Models (SAMs (Kerner and Klenov, 2006)), and the Average Space Gap cellular automaton Model (ASGM (Tian et al., 2012a, 2012b)). However, some of these models have been criticized by proponents of three-phase theory. BLM has been criticized because its congested patterns are inconsistent with the empirical findings (Kerner et al., 2002). SAMs are not able to reproduce the observed local synchronized patterns (LSPs) as well as some of empirical features of synchronized flow between wide moving jams within general patterns (GPs) (Kerner and Klenov, 2006). However, these criticisms cannot be applied to the ASGM which can describe the LSPs and GPs very well.

Although this paper is motivated by the inconsistency between fundamental diagram approach and three-phase theory, the purpose is not to discuss their controversies. Instead, this paper aims to describe the driver behavior by a cellular automaton containing explicit oscillations around the steady-state in the noiseless limit thereby reproducing the major observational aspects of three-phase theory. Moreover, empirical calibration and validation results in a higher accuracy than that of previously investigated models (Brockfeld et al., 2005; Wagner et al., 2010). To these ends, Section 2 analyses the US-101 trajectory datasets on a single freeway lane, away from lane changes and the influence of bottlenecks. Section 3 proposes a cellular automaton model that incorporates this assumption. Empirical findings of three-phase theory are simulated and discussed in Section 4. Section 5 is devoted to calibrating and validating the model to the I-80 detector data. The concluding Section 6 gives a summary and a discussion.

## 2. Empirical data analysis

The essential distinction between the fundamental diagram approach and the three-phase theory is whether the fundamental diagram exists. Three-phase theory supposes the existence of a two-dimensional region in the density-flow plane or, equivalently, in the gap-speed plane. Drivers can make an arbitrary choice in the space gap within a certain region. The fundamental diagram approach assumes the existence of a unique gap-speed relationship. For example, the Intelligent Driver Model (IDM) presumes the following relationship (Treiber et al., 2000):

$$s^*(v, \Delta v) = s_0 + vT + \frac{v\Delta v}{2\sqrt{ab}} \tag{1}$$

Here, $s^*$ is the desired gap. The meaning and typical values of the IDM parameters are shown in Tab.1. The relationship between the space gap $s$ and speed $v$ in the steady state is $s=s_0+vT$, which is also assumed in other car following models such as Newell's model (Newell, 2002). Thus, the validation of the relation $s=s_0+vT$ contributes to resolving these controversies between the fundamental diagram approach and three-phase theory. However, it is impossible to validate this relationship by the real traffic data directly, since real traffic flow is always away from the steady state, at least, to some extent. Nevertheless, Equation (1) indicates that we can validate this relationship, if the speed difference between vehicles is approximately zero. Therefore, in order to exclude the influence of the speed difference $\Delta v$, we only analyze the empirical data with $|\Delta v| \leq \Delta v_c$. The value of $\Delta v_c$ should be able to neglect the influence of $\Delta v$ and, nevertheless, make the empirical sample size large enough. Assuming $|\Delta v| \leq \Delta v_c$, we have $s^* \approx s_0 + vT$. A suitable value for $\Delta v_c$ is 0.1 $m/s$.

**Table 1**
Model Parameters and values of IDM (Treiber and Kesting, 2013).

| Parameter (unit) | Typical value | Reasonable range |
|---|---|---|
| Safe time gap $T(s)$ | 1.5 | 0.9~3 |
| Minimum gap $s_0(m)$ | 2 | 1~5 |
| Acceleration $a$ ($m/s^2$) | 1.4 | 0.3~3 |
| Comfortable deceleration $b$ ($m/s^2$) | 2.0 | 0.5~3 |

Next, the US-101 trajectory data sets of the Next Generation Simulation Community (NGSIM, 2006) are applied to validate $s=s_0+vT$. These data were collected on a 640$m$ segment on the south-bound direction of US 101 (Hollywood Freeway) in Los Angeles, California on June 15th, 2005. The data were detected from 7:50 a.m. to 8:05 a.m., 8:05 a.m. to 8:20 a.m., and 8:20 a.m. to 8:35 a.m. During the data collection period, the California Highway Patrol's Computer Aided Dispatch (CHP CAD) system was monitored. No traffic incidents were recorded during the morning of June 15th on US-101 within the study area or on any upstream/downstream sections likely influencing traffic in the study area. Fig.1 provides a schematic illustration of the location for the vehicle trajectory datasets. There are five mainline lines throughout the section, and an auxiliary lane is present through a portion of the corridor between the on-ramp and off-ramp. In order to minimize the impact of bottlenecks on traffic flow, only the leftmost lane is analyzed. The following criteria are used to filter suitable trajectories from the empirical data:

1. The vehicle's leading car could not change lanes during the whole period,
2. the absolute value of the vehicle speed difference to the leader is smaller than 0.1$m/s$,
3. there are at least 100 data points for the selected vehicle, i.e. a trajectory duration of 10 seconds or more,
4. the space headway should be shorter than 76$m$ (250$ft$). Otherwise, it is likely that there is no interaction (Bham and Benekohal, 2004).

After applying these criteria, 323 out of 1226 vehicle trajectories were included in the analysis, see Tab.2. The optimal values of $T$ and $s_0$ for each vehicle were estimated by the linear regression analysis. Some results are shown in Figs.2 and 3. The linear relationship between the actual space gap *s* and speed *v* are identified when the speed difference approaches zero. Figs.2 and 3 also illustrate that vehicles will not keep uniform speeds although the actual space gap *s* approximates desired space gap $s^*$ calculated by $s^* = s_0 + vT$, and speed difference *Δv* approximates zero. This means that, even if the stimulus vanish, i.e., $\Delta v \to 0$ and $s \to s^*$, the vehicles accelerate or decelerate most of the time. While this could possibly be explained by the drivers' anticipation or finite reaction time, we make the following assumptions to explain this phenomenon from another perspective: *In homogeneous congested traffic flow and in the noiseless limit, the space gap will oscillate around the (speed dependent) desired value rather than maintain this value.* In this perspective, no steady states exist in the homogeneous congested traffic flow. This is different from the fundamental diagram approach or the three-phase theory, which admit the existence of definite steady states or a multitude of steady states, respectively. In this perspective, even if the actual space gap equals the desired space gap and the speed difference is approximately zero, vehicles will accelerate or decelerate in an oscillatory way. It should be noted that many car following models considered very small fluctuations around the desired space gap, i.e. acceleration noise. However, it is impossible that the acceleration noise is responsible for the high observed values of accelerations and decelerations depicted in Fig.3.

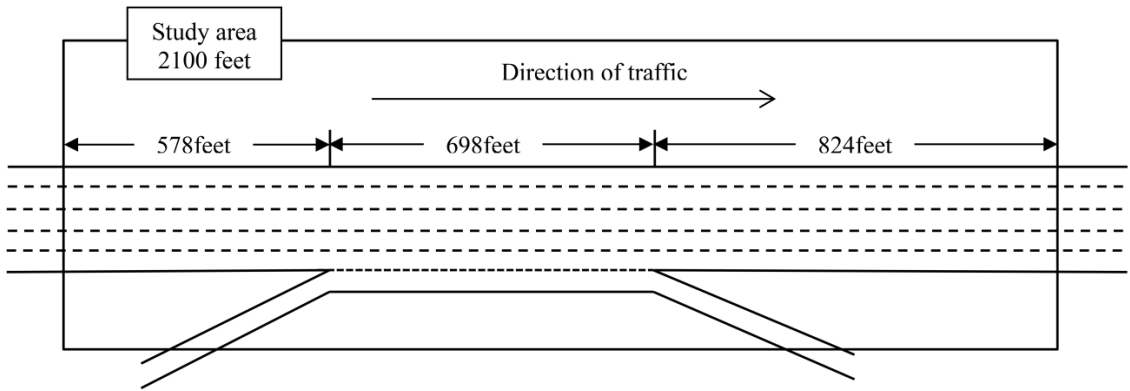

**Fig. 1.** The sketch of US-101 study area.

**Table 2**
Overview of F-Tests of the linear relationship between space gap and speed.

| Time interval | Number of simples available before filtering | Number of simples available before testing for significance | Number of simples available after testing for significance | |
| --- | --- | --- | --- | --- |
| | | | $s_0$ | lower 95% confidence bounds of $s_0$ greater than zero |
| 07:50am-08:05am | 430 | 84 | 84 | 74 |
| 08:05am-08:20am | 410 | 98 | 97 | 78 |
| 08:20am-08:35am | 386 | 141 | 139 | 113 |
| Total | 1226 | 323 | 320 | 265 |

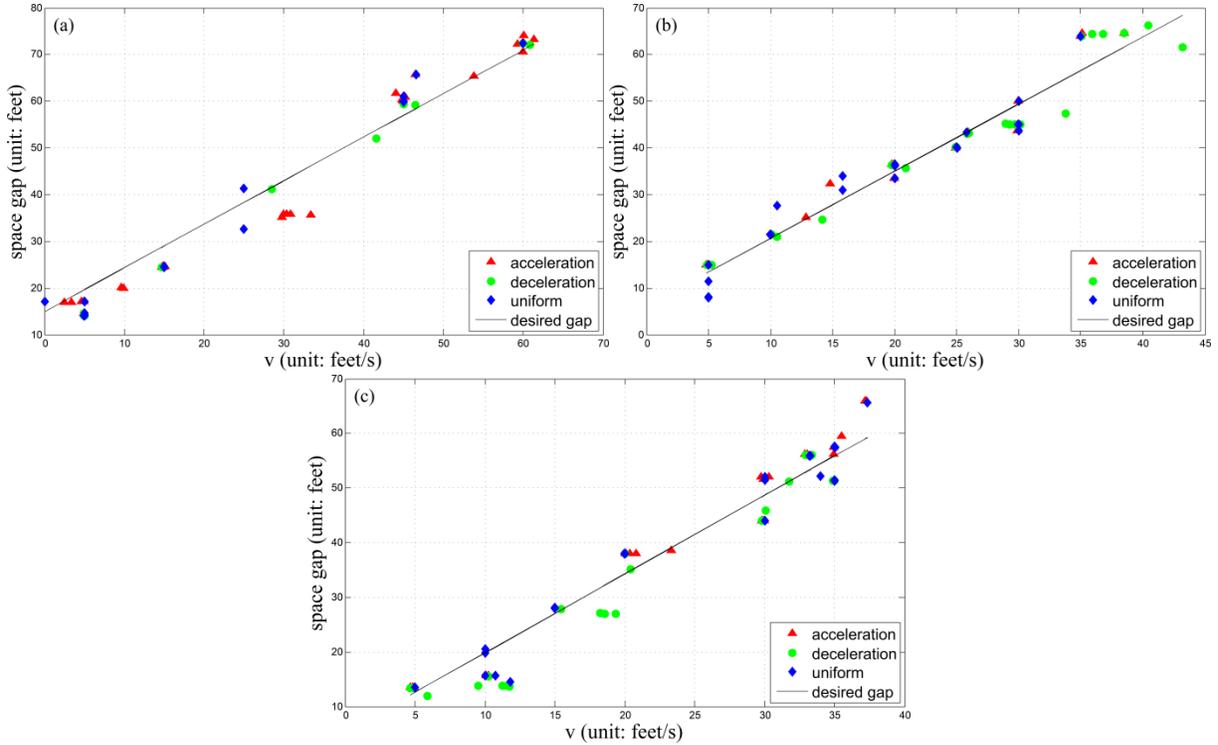

**Fig.2.** Linear regression analysis for the space gap as a function of the speed. (a-c) are single-vehicle data taken from different detecting time intervals 7:50 a.m. to 8:05 a.m., 8:05 a.m. to 8:20 a.m., and 8:20 a.m. to 8:35 a.m., respectively. Speed, acceleration and space gap are the instantaneous speed, acceleration and space gap respectively, which are taken from the raw record data without any processing. If the acceleration or deceleration is less than $0.1 m/s^2$, we assume a steady-state situation.

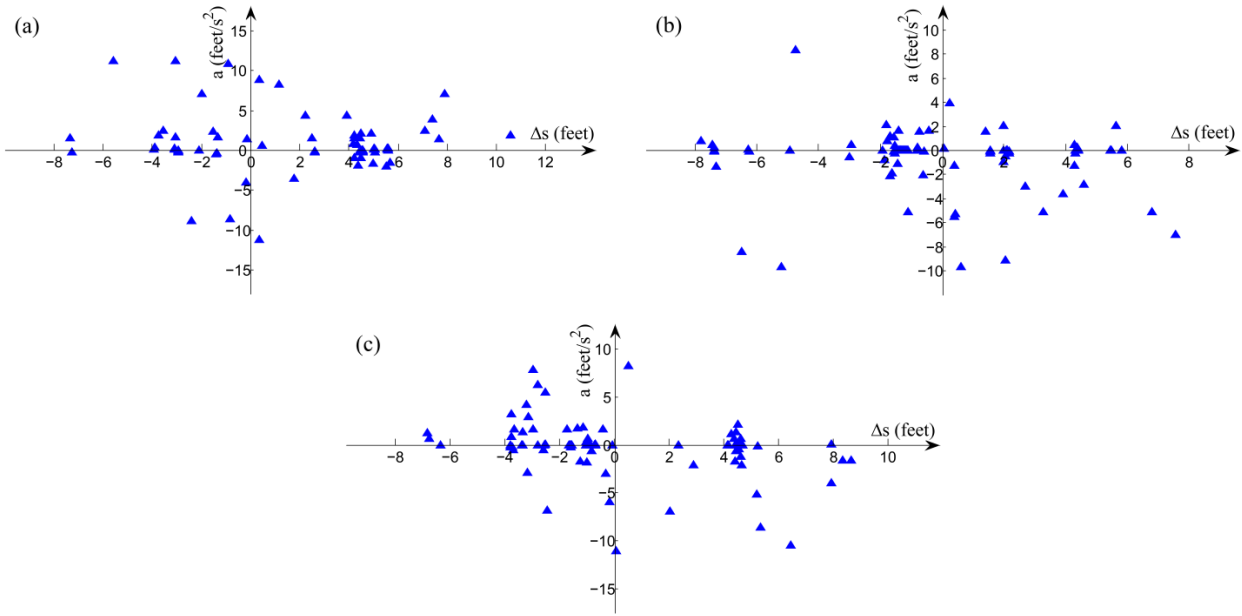

**Fig.3.** The accelerations corresponding to Fig. 2(a-c), respectively. $\Delta s=s-s_0-vT$ denotes the deviation from the steady-state gap.

F-Tests for significance of the linear relationship at a level of 5% are also performed. The results are shown in Tab.2 and Fig.4. Tab. 2 shows that almost all data indicate the existence of the linear relationship, although some values of $s_0$ are less than zero. If the data with the lower 95% confidence bound of $s_0$ less than zero are excluded, there remain still 265 samples (82%).

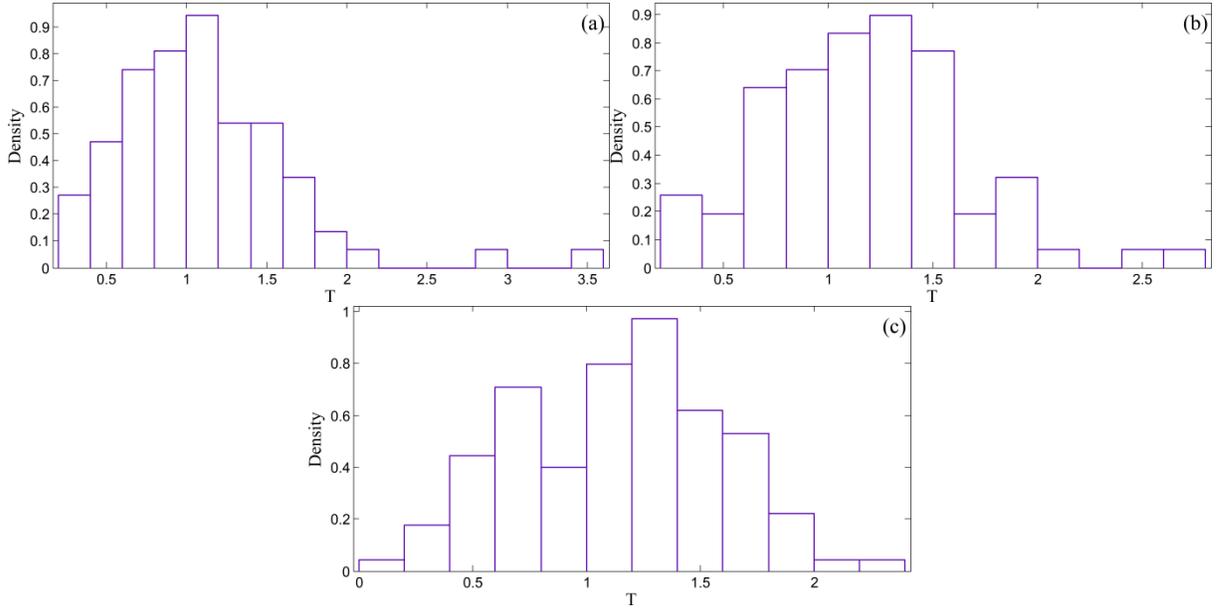

**Fig.4.** Histograms of the safe time gaps. (a-c) correspond to the detecting time intervals 7:50 a.m. to 8:05 a.m., 8:05 a.m. to 8:20 a.m., and 8:20 a.m. to 8:35 a.m., respectively.

## 3. The new model

We propose a new cellular automaton model based on the above assumption, i.e., vehicles' space gap will oscillate around the desired space gap in the homogeneous congested traffic flow in the noiseless limit. The main mechanisms incorporating this assumption are embodied in the randomization process of vehicles. The parallel-update rules are as follows.

1. Determination of the randomization parameter $p_n(t+1)$ and deceleration extent $\Delta v$:

$$p_n(t+1) = \begin{cases} p_a: & \text{if } d_n^{eff}(t) < d_n^*(t) \\ p_b: & \text{if } v_n(t) = 0 \text{ and } t_n^{st}(t) \geq t_c \\ p_c: & \text{in all other cases} \end{cases} \quad (2)$$

$$\Delta v(t+1) = \begin{cases} b_{defens}: & \text{if } d_n^{eff}(t) < d_n^*(t) \\ 1: & \text{in all other cases} \end{cases} \quad (3)$$

2. Acceleration:

$v_n(t+1) = \min(v_n(t)+1, v_{max})$

3. Deceleration:

$v_n(t+1) = \min(d_n^{eff}(t), v_n(t+1))$

4. Randomization with probability $p_n(t+1)$:

if $(rand() < p_n(t+1))$ then $v_n(t+1) = \max(v_n(t+1) - \Delta v(t+1), 0)$

5. Determination of $t_n^{st}(t+1)$:

if $(v_n(t+1) = 0)$ then $t_n^{st}(t+1) = t_n^{st}(t) + 1$

if $(v_n(t+1) > 0)$ then $t_n^{st}(t+1) = 0$

6. Car motion:
$$x_n(t+1) = x_n(t) + v_n(t+1)$$

Here, $d_n(t)$ is the space gap between vehicle $n$ and its preceding vehicle $n+1$, $d_n(t) = x_{n+1}(t) - x_n(t) - L_{veh}$, $x_n(t)$ is the position of vehicle $n$ (here vehicle $n+1$ precedes vehicle $n$), and $L_{veh}$ is the length of the vehicle. Furthermore, $v_n(t)$ is the speed of the vehicle $n$, $v_{max}$ is the maximum speed, $d_n^*(t) = Tv_n(t)$ is the effective desired space gap between vehicle $n$ and $n+1$, and $T$ is the effective safe time gap between vehicle $n$ and $n+1$ at the steady state, and $d_n^{eff}(t) = d_n(t) + max(v_{anti}(t) - g_{safety}, 0)$ is the effective gap. In this definition, $v_{anti} = min(d_{n+1}(t), v_{n+1}(t)+1, v_{max})$ is the expected speed of the preceding vehicle in the next time step, and $g_{safety}$ the parameter to control the effectiveness of the anticipation. Accidents are avoided only if the constraint $g_{safety} \geq b_{defens}$ is satisfied. The speed anticipation effect is considered in order to reproduce the real time headway distribution, which has a cut off at the small time headway less than one second (Neubert et al., 1999). $t_n^{st}(t)$ denotes the time since the last stop for standing vehicles, while $t_n^{st}(t) = 0$ for moving vehicles.

The basis of the new model is the rule set of the NaSch model with randomization parameter $p_c$ to which a slow-to-start rule and the effective desired space gap $d_n^*(t)$ have been added. The slow-to-start effect is characterized by an increase of the randomization parameter from $p_c$ to $p_b$ ($> p_c$), which is the element to realize the transition from synchronized flow to wide moving jams. The new model assumes the driver tends to keep the effective gap no smaller than $d_n^*(t)$, otherwise the driver will become defensive. The actual behavioral change is characterized by increasing the spontaneous braking probability from $p_c$ to $p_a$. Moreover, the associated deceleration will change from 1 to $b_{defens}$ ($\geq 1$). This effect is the factor to reproduce the transition from free flow to synchronized flow in the new model.

In the following, the steady states of the new model are analyzed in the unperturbed, noiseless limit. For microscopic traffic flow models, the steady state requires that the model parameters are the same for all drivers and vehicles. In that case, the steady state is characterized by the following two conditions (Treiber and Kesting, 2013):

*1) Homogeneous traffic*: All vehicles move at the same speed and keep the same gap behind their respective leaders.

*2) No accelerations*: all vehicles keep a constant speed.

Since the mechanisms associated with the hypothetical congested steady state analysis are all embodied in the randomization process, the noiseless limit should be taken as $p_a = 1, p_c = 0, p_b = 0$ or $p_a = 1, p_c = 1, p_b = 1$. However, all vehicles will keep a constant speed no matter how far distance between vehicles in the latter case, which is obviously unrealistic. Thus, we consider the former. According to the model rules, if $d^{eff}/T \geq v_{max}$, all vehicles will move with $v_{max}$; if $d^{eff}/T < v_{max}$, all vehicles' speed will take turns to change simultaneously over time between $max(v - b_{defens}, 0)$ and $v$, where $v \in [d^{eff}/T, min(v_{max}, d^{eff})]$ and $max(v-1, 0) < d^{eff}/T$. This means that there are no steady states of congested traffic in the new model. Instead, the space gaps oscillate around the desired gap, i.e., fluctuations are caused by the internal randomness of the drivers, not (only) by the driver heterogeneity, which is consistent with the empirical findings by Wagner (2012). Thus, this model is named as the cellular automaton model with non-hypothetical congested steady state (NH model). Finally, we note that by modifying the acceleration rule of the NH model as follows, we obtain a cellular automaton model within the fundamental diagram approach:

$$\text{if } d_n^{eff}(t) > d_n^*(t) \text{ then}: v_n(t+1) = \min(v_n(t)+1, v_{max}) \tag{4}$$

## 4 Simulation investigation

In this section, simulations are carried out on a road of length $L_{road} = 1000L_{cell}$. Both the cell length and vehicle length are set as 7.5$m$. One time step corresponds to 1$s$ in reality. During the simulations, the first 50000 time steps are discarded to let the transients die out. The parameters are shown in Tab.3.

**Table 3**
Model parameters of NH model.

| Parameters | $L_{cell}$ | $L_{veh}$ | $v_{max}$ | $T$ | $b_{defens}$ | $p_a$ | $p_b$ | $p_c$ | $g_{safety}$ | $t_c$ |
|---|---|---|---|---|---|---|---|---|---|---|
| Units | $m$ | $L_{cell}$ | $L_{cell}/s$ | $s$ | $L_{cell}/s^2$ | - | - | - | $L_{cell}$ | $s$ |
| Value | 7.5 | 1 | 5 | 1.8 | 1 | 0.95 | 0.55 | 0.1 | 2 | 8 |

*4.1. Periodic boundary condition*

Periodic boundary conditions reflect a ring road, i.e., the first vehicle $N$ is connected to the last vehicle 1 by the boundary conditions

$$v_{N+1} = v_1, \ d_{N+1} = x_1 + L_{road} - x_n - L_{veh} \tag{5}$$

Figure 5 shows the macroscopic flow-density diagram by plotting the time-averaged flux $f$ (vehicles per hour) at a given location over the global density $K$. Since the number $N$ of vehicles and thus the global density $K = N/L_{road}$ is fixed, each data point corresponds to an individual simulation.

One can see that there are two branches in the density region $K_1 < K < K_3$: the upper branch is obtained from the initially homogeneous distribution of traffic whereas the lower starts from a wide moving jam. Therefore, three traffic phases and two first order transitions (the transitions from free flow to synchronized flow (F→S) and from synchronized flow to wide moving jams (S→J)) are clearly distinguished, exhibiting a typical double Z-characteristic structure predicted by the three-phase traffic flow theory. Moreover, when the density increases ($K_2 < K < K_3$), the flux begins to decrease and the synchronized flow starts to emerge in the free flow when the initially state is homogeneous traffic (Fig.6(a),(b)). While the initially wide moving jams traffic will evolve to the state that wide moving jams and free flow coexist (Fig.6(c),(d)).

Next, we distinguish the synchronized flow and the wide moving jam phases with single traffic data by the flow interruption effect (see Appendix). We obtained the data by setting a virtual detector on the road. Fig.7 shows the sketch of virtual detector on the road. The detector measures the numbers and speeds of the vehicles that pass it in the aggregation time interval 60$s$. Since the state of each cell is known, we can tell whether the cell is occupied by the vehicle. Therefore, if the detector location is within the wide moving jam, there is no vehicle passing it. Both the number and speed of vehicles are zero. It is different from the real detector, which cannot measure speed if the number of vehicles is zero because one cannot tell apart stopped traffic from an empty road. The interruption effect can be clearly identified in the Figs.8(a),(b). Before and after the wide moving jam has passed the detector, many vehicles traversed the detector. But within the jam, no vehicles traversed the detector, and the speed within the jam is zero. This means that the traffic flow is discontinuous within the moving jam, i.e., this moving jam is associated with the wide moving jam phase. The flow interruption does not occur in the Figs.8(c),(d). Thus, this is associated with the synchronized flow phase.

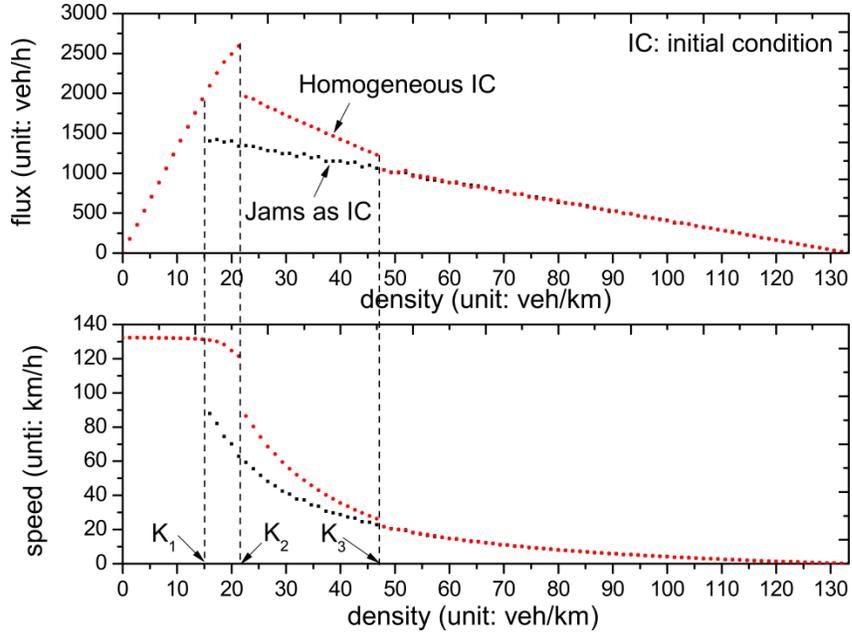

**Fig.5.** Flow-density diagram of NH model. "Homogeneous IC" represents the initially homogenous distribution of traffic, and "Jams as IC" represents the initially wide moving jam distribution of traffic. The upper lines between $K_2$ and $K_3$ in the flux-density and speed-density plots, respectively, correspond to a mixture of coexisting free traffic and synchronized phases, while the lower line between $K_1$ and $K_3$ corresponds to a coexistence of free traffic and jams.

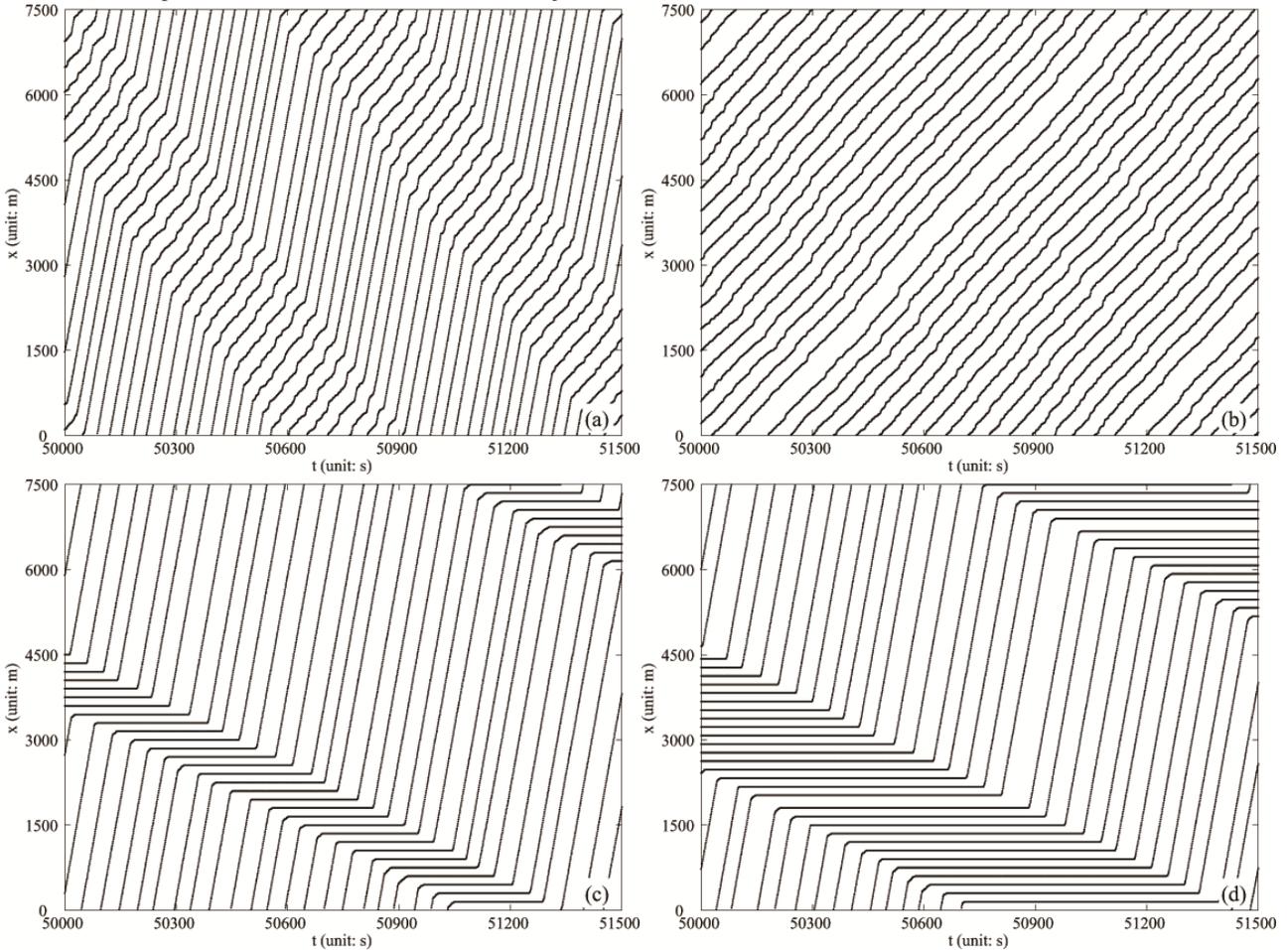

**Fig.6.** Trajectories of every 20<sup>th</sup> vehicle of the NH model, (a) *k*=27, (b) *k*=47, (c) *k*=27, (d) *k*=47 (unit: *veh/km*). (a),(b) Starting from homogeneous initial state. (c),(d) Starting from a wide moving jam initial state. The horizontal direction (from left to right) is time and the vertical direction (from down to up) is space.

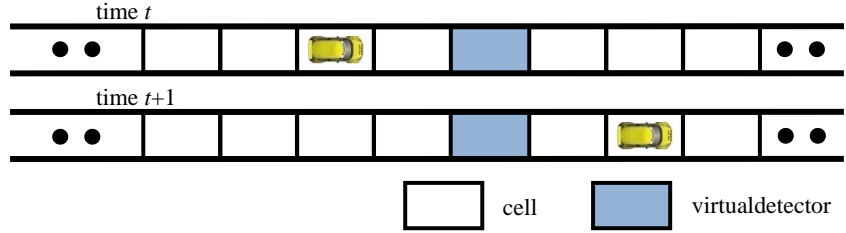

**Fig.7.** The sketch of a virtual detector.

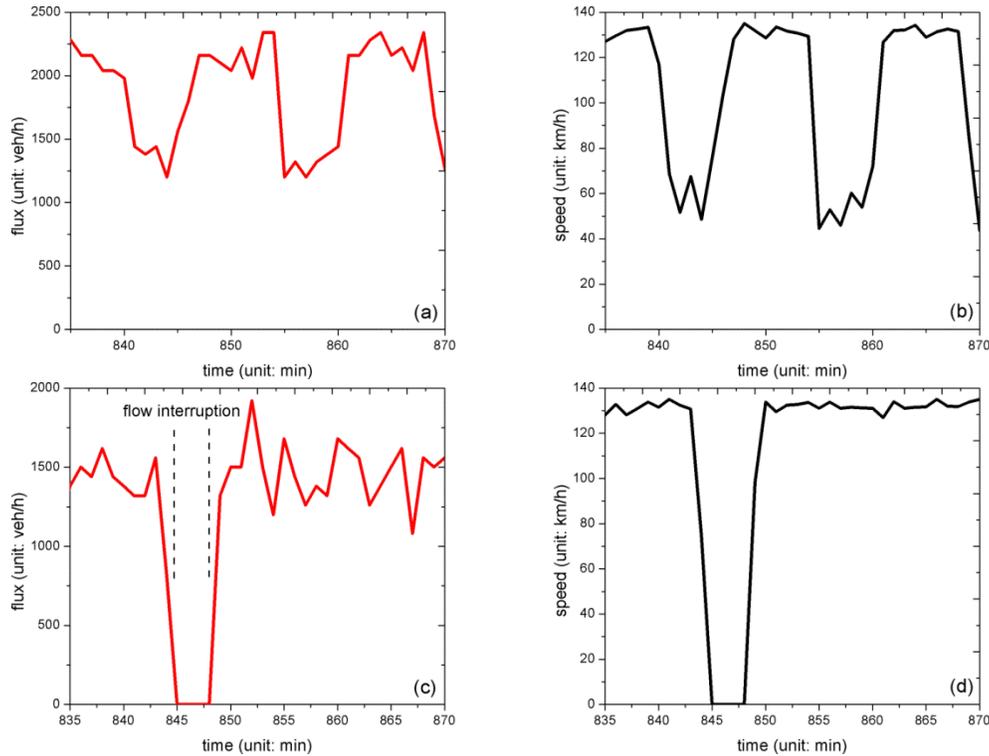

**Fig.8.** (a), (c) The 1 min average flux. (b),(d) The 1 min average speed. (a),(b) stating from homogenous initial state responding to Fig.6(a). (c),(d) starting from a wide moving jam initial state responding to Fig.6(c).

*4.2. Open boundary condition*

The traffic patterns that emerge near an on-ramp are studied under open boundary condition. The vehicles drive from left to right. The left-most cell corresponds to $x=1$. The position of the left-most vehicle is $x_{last}$ and that of the right-most vehicle is $x_{lead}$. At each time step, if $x_{last}>v_{max}$, a new vehicle with speed $v_{max}$ will be injected to the position $\min(x_{last}-v_{max}, v_{max})$ with probability $q_{in}/3600$ where $q_{in}$ is the traffic flow entering the main road in units of vehicles per hour. At the right boundary, the leading vehicle moves without any hindrance. If $x_{lead}>L_{road}$, the leading vehicle will be removed and the following vehicle becomes the leader.

We adopt a simple method to model the on-ramp, which is similar to that of Treiber et al. (2006). Assuming the position of the on-ramp is $x_{on}$, a region $[x_{on}, x_{on}+L_{ramp}]$ is selected as the inserting area of the vehicle from on-ramp. At each time step, we find out the longest gap in this region. If the gap is large enough for a vehicle, then a new vehicle will be inserted at the cell in the middle of the gap with probability $q_{in}/3600$ and $q_{on}$ is the traffic flow from the on-ramp. The speed of the inserted vehicle is set as the speed of its preceding vehicle, and the stop time is set to zero. The parameters are set as $x_{on}= 0.8L_{road}$ and $L_{ramp}=10L_{cell}$.

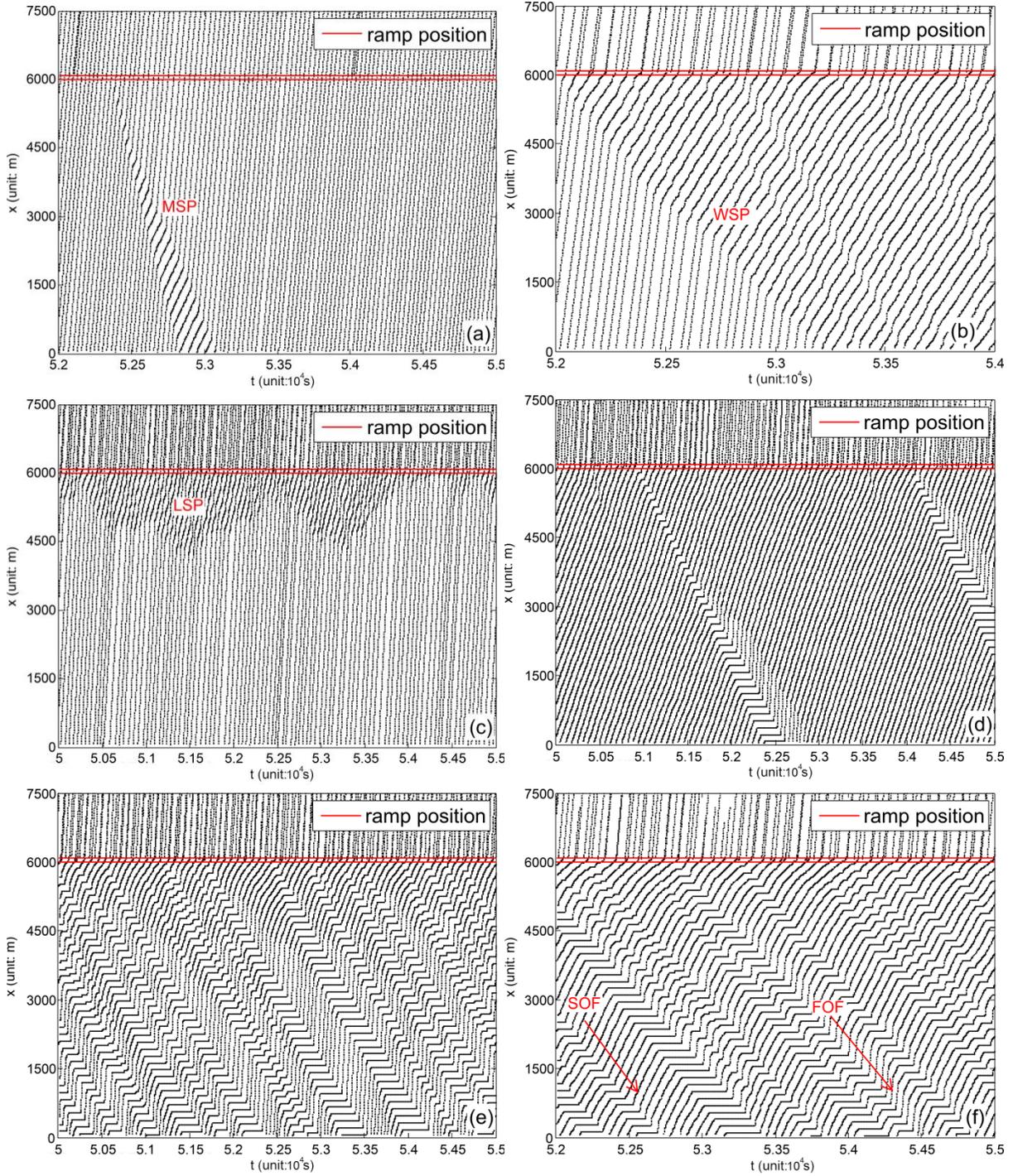

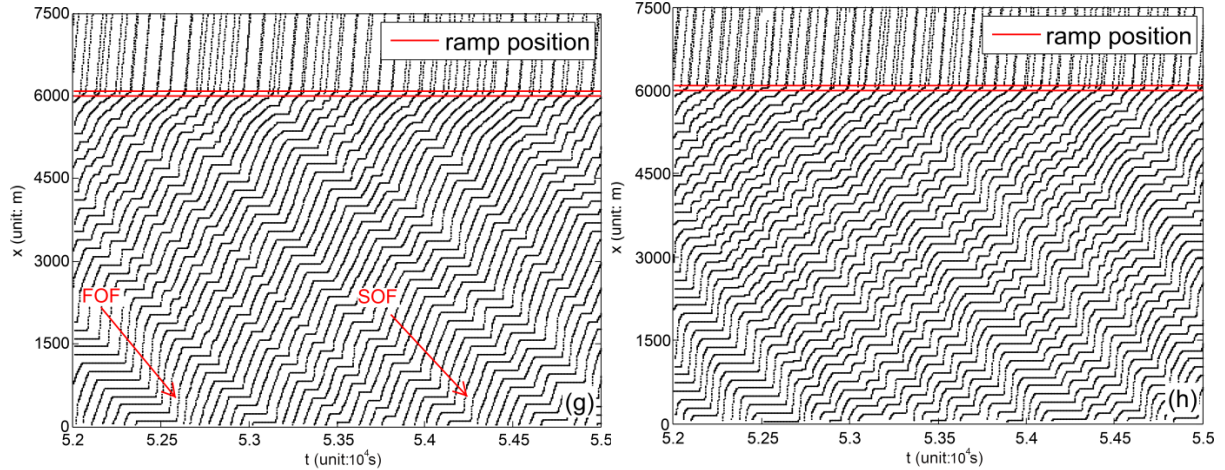

**Fig.9.** Trajectories of every 20$^{th}$ vehicle of the NH model. (a)$q_{in}$=2339, $q_{on}$ =19 (MSP), (b)$q_{in}$=1728, $q_{on}$ =968 (WSP), (c)$q_{in}$=1440, $q_{on}$ =823 (LSP), (d) $q_{in}$=1134, $q_{on}$ =1123 (DGP), (e) $q_{in}$=920, $q_{on}$ =1304 (GP), (f) $q_{in}$=931, $q_{on}$ =1304 (GP), (g) $q_{in}$=933, $q_{on}$ = 1011 (GP), (h)$q_{in}$=907, $q_{on}$ =1410 (GP) (unit: veh/h). The horizontal direction (from left to right) is time and the vertical direction (from down to up) isspace. (f) $p_b$=0.5, (g) $p_b$=0.55, $T$=1.6, $g_{safety}$ =$b_{defens}$=2, (h) the acceleration rule is according to equation (4). 'SOF' and 'FOF' represent the synchronized outflow and free outflow of wide moving jams, respectively.

In Fig.9(a), the spatial-temporal features of the congested pattern named moving synchronized flow (MSP) reproduced are shown (see the empirical figure 7.6 in Kerner (2009)). In this pattern, synchronized traffic flow spontaneously emerges in the free flow. Fig.9(b) exhibits the widening synchronized flow (WSP, see the empirical figure 7.4 in Kerner (2009)). For this pattern, wide moving jams do not emerge in synchronized flow. The downstream front of WSP is fixed at the on-ramp and the upstream front of WSP propagates upstream continuously over time. In Fig.9(c), both the downstream and the upstream front of synchronized flow are fixed at the on-ramp, thus, it belongs to the local synchronized pattern (LSP). Moreover, the width of LSP in the longitudinal direction changes over time, which is in accordance with empirical observations (see the empirical figure 7.2 in Kerner (2009)). Fig.9(d) shows the dissolving General Patterns (DGP) in which just one wide moving jam emerges in the synchronized flow. Fig.9(e) shows the spatial-temporal features of General Pattern (GP). Only free outflow exists in the downstream of wide moving jams in GP, which has been criticized by the three-phase theory. However, it can be easily improved if we decrease the slow-to-start probability $p_b$ or adjust the values of $T$ and $b_{defens}$, see Fig.9(f),(g). Thus, all the above simulation results arewell consistent with the well-known results of the three-phase traffic theory. However, if the acceleration rule is revised as equation (4), the synchronized outflow cannot be reproduced any more.

In Fig. 9(a) and (d), one could obtain the propagation velocity of the downstream MSP front is nearly −26.8*km/h* and the propagation velocity of the downstream jam front is nearly −13*km/h* which is about half that of the downstream MSP front. This is better than in most three-phase models which often have propagation velocities as negative as *-40km/h* or even more negative.

## 5. Empirical validation

In order to validate NH model, comparing the simulating data with the empirical data is necessary. The datasets presented by NGSIM are from double loop detectors between Powell Street and Gilman Avenue on Interstate 80 (I-80) in Emeryville, California, see Fig.10. The I-80 is a five-lane freeway. The data were collected through Freeway Performance Measurement System (PeMS) project, which was conducted by the Department of Electrical Engineering and Computer Sciences at the University of California, at Berkeley, with the cooperation of California Department of Transportation. Available data from six detector stations (Stations 1, 3, 4, 5, 6 and 7) were provided in this data set.

Each detector station contains two detectors per lane. This data set provides 30-second processed, loop detector data. Speed (unit: *feet/s*), volume (unit: *number*) and occupancy (unit: *percentage*) at each detector for the 30-second time step are presented at each detector in each lane. The NH model will be calibrated with the data of Thursday, 07 April 2005 and then validated with the data of other five days.

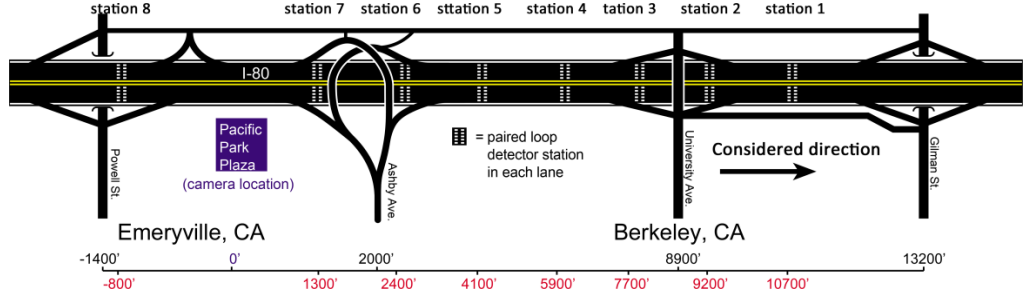

**Fig.10.** The sketch of I-80 near Berkeley.

As in the investigation of Brockfeld et al. (2005), we average the speed and flux data over the *N* lanes to obtain effective single lane speeds $v_{ave}^i$ and fluxes $f_{ave}^i$ as follows:

$$v_{ave}^i = \sum_{j=1}^{N} w_{ij} v_{ij}^{emp}, \quad w_{ij} = \frac{f_{ij}^{emp}}{\sum_{j'=1}^{N} f_{ij'}^{emp}}, \tag{5}$$

$$f_{ave}^i = \frac{1}{N} \sum_{j=1}^{N} f_{ij}^{emp} \tag{6}$$

where $f_{ij}^{emp}$ and $v_{ij}^{emp}$ are the empirical flux and speed of lane *j* detected by the station *i*. Empirical fluxes $f_{ij}^{emp} = n_{ij}^{emp}/30$ (unit: *veh/s*) are defined by dividing the number $n_{ij}^{emp}$ of vehicles passing in each time interval 30s over station *i* on lane *j* by this interval. Since NH model is a single-lane model, this avoids the additional complexity of lane-change rules needed to perform multi-lane simulation. Free traffic and synchronized flow can be identified in Fig.11 (a-e), which shows the time series of speed for each lane at Station 5 on Thursday, 07 April 2005. Fig.11(f) is the average speed ($v_{ave}^5$) data on the station 5, derived from Equation (5). Comparing with Fig.11(a-e), Fig.11(f) is capable of reflecting the same traffic dynamics around Station 5. Thus, this method is useful to test whether the NH model can describe the realtraffic flow patterns.

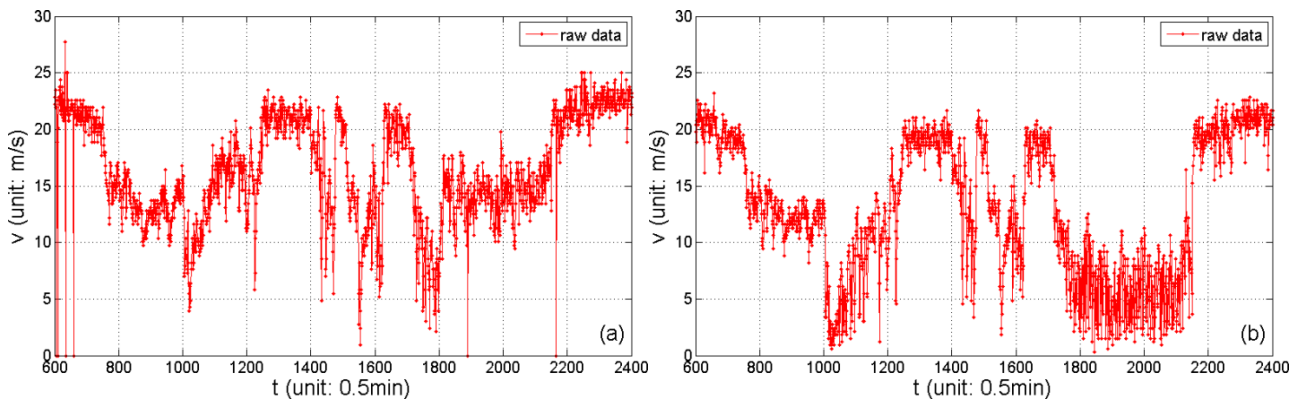

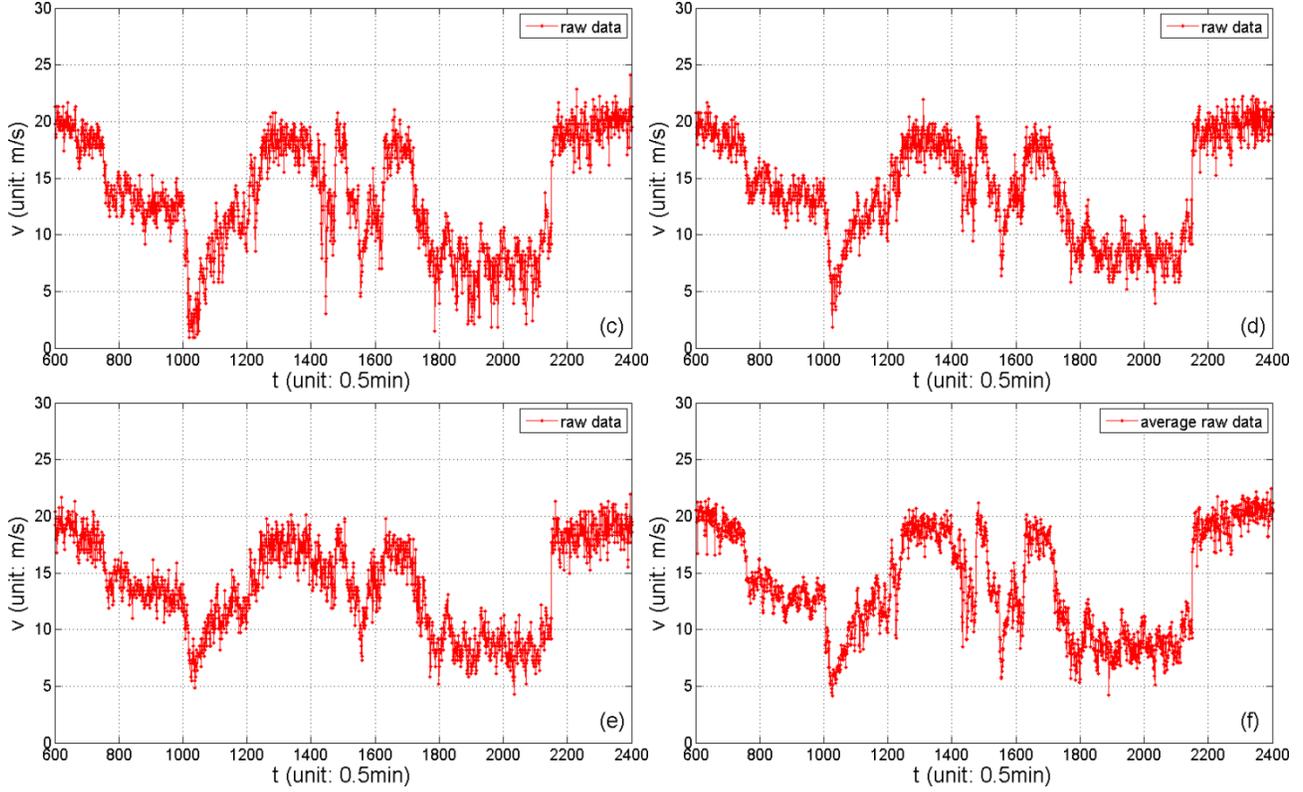

**Fig.11.** Real Time series of speed at Station 5 on Thu, 07 Apr 2005. (a-e) correspond to the real speed series lane 1 to lane 5 detected by station 5. (f) is the real lane average speed series calculated by Equation (5).

*5.1. Simulation setup*

The vehicles are generated at inflow Station 6. To insert new vehicles, the in-flowing open boundary condition in section 4.2 is applied. The position of station 6 is set at the left-most cell. At each time step, if $x_{last} > v_{max}$, a new vehicle with speed $v_{max}$ will be injected to the position min($x_{last}$-$v_{max}$, $v_{max}$) with probability $q_{in} = f_{ave}^6$.

To regulate the outflow, the limit speed region closely downstream of Station 4 is defined, which starts from the position of Station 4, and ends with the length $L_{sl}$=50$m$, i.e., the speed limit region is located at [1067, 1117]$m$. In the speed limit region, a dynamic speed limit is applied, which equals the speed $v_{ave}^4$ that Station 4 measured. Since the CA models require an integer value for the speed limit, we converted the value to $\lfloor v_{ave}^4 / L_{cell} + 1 \rfloor$, where $\lfloor x \rfloor$ denotes the maximum integer that is not bigger than $x$. The simulation ends at 1117$m$.

*5.2. Goodness-of-fit measures*

Theil's inequality coefficient (U) is applied to measure the performance (Brockfeld et al., 2005; Ahmed, 1999), which is defined as follows:

$$U = \frac{\sqrt{\frac{1}{N}\sum_i (v_{ave}^{5i} - v_{simu}^{5i})^2}}{\sqrt{\frac{1}{N}\sum_i (v_{ave}^{5i})^2} + \sqrt{\frac{1}{N}\sum_i (v_{simu}^{5i})^2}} \quad (6)$$

where $v_{simu}^{5i}$ is the *i*th speed of the simulation data at station 5. $v_{ave}^{5i}$ is the *i*th lane average speed of the empirical data calculated by Equation (5) at station 5. N is the number of the data points. The value of $U$ is always in the range between 0 and 1 with $U=0$ implying a perfect fitness. Related to Theil's U, the bias ($U^M$) and the variance ($U^S$) are often applied:

$$U^M = \frac{(\mu_{v_{ave}^5} - \mu_{v_{simu}^5})^2}{\frac{1}{N}\sum_i (v_{ave}^{5i} - v_{simu}^{5i})^2} \quad (7)$$

$$U^S = \frac{(\sigma_{v_{ave}^5} - \sigma_{v_{simu}^5})^2}{\frac{1}{N}\sum_i (v_{ave}^{5i} - v_{simu}^{5i})^2} \quad (8)$$

where $\mu_{v_{ave}^5}$, $\mu_{v_{simu}^5}$, $\sigma_{v_{ave}^5}$ and $\sigma_{v_{simu}^5}$ are the means and standard deviations of the empirical and the simulated speed series at station 5, respectively. The bias proportion ($U^M$) reflects the systematic error. The variance proportion ($U^S$) indicates how well the fluctuation in the original data is replicated by the simulation. Therefore, lower values (close to zero) of $U^M$ and $U^S$ are desired.

*5.3. Calibration*

In contrast to previous works (Brockfeld et al., 2005; Wagner et al., 2010), the automated calibration methods are not used, since they did not always converge. Moreover, once the cell length $L_{cell}$ is determined, the parameters of the NH model can be calibrated in a straightforward way. The maximum speed $v_{max}$ can be taken as the maximum of the empirical data if there are periods of free flow (otherwise, it cannot be estimated). Through the average free flow speed $v_{free}^{ave}$, the randomization probability $p_c$ is calculated by $v_{free}^{ave} = v_{max} - p_c$. The randomization probability $p_b$ can be estimated by $v_g \approx -(1-p_b)/k_{max}$ if the downstream propagation speed of the wide moving jam $v_g$ is determined, where $k_{max}$ is the density inside the jam. Rehborn et al. (2011) have discussed several methods to measure $v_g$ and found it inside the interval [-18, -10] km/h while Treiber et al. (2010) give the interval [-20,-15] km/h. The parameter $t_c$ affects the emergence of wide moving jams. Given a fixed probability $p_b$, a larger $t_c$ increases the probability that congested traffic is in the synchronized state, i.e., it reduces the probability of an S→J transition (Jiang and Wu, 2003). The parameter $g_{safety}$ is not smaller than the deceleration $b_{defens}$ to keep safety. It can be adjusted after other parameters are determined. Furthermore, we found it is best not to unnecessarily change the value of randomization probability $p_a$. Since the vehicle length $L_{veh}$ is about 7.5m, it can be identified after $L_{cell}$ is given. Thus, only $L_{cell}$, $T$ and $b_{defens}$ are left to be adjusted, which influence the state of the synchronized flow. The trial and error method is adopted to determine their values.

During the simulations, we found that a smaller cell length $L_{cell}$ is needed to make the simulation data more consistent with the empirical data and $L_{cell}$=1m is good enough to obtain satisfactory results (cf. Table 4). Since the maximum speed measured by the detectors is around 20 *m/s*, we set $v_{max}$=20$L_{cell}$/s. As the vehicle types and lengths are unknown, only one type of vehicle of length $L_{veh}$=7$L_{cell}$ = 7*m* is assumed. The wide moving jams have not been detected, so $p_b$ and $t_c$ will not be changed from the values of Table 3. Figures 12(b) and (c) visualize the effect of the calibration procedure. While a simulation with the un-calibrated values (Table 3) results in a poor fit (Fig. 12(b)), we obtain a good agreement after calibration (Fig. 12(c)). Specifically, using the un-calibrated values, we always obtain free flow which means the synchronized flow in front is underestimated. Thus, we need to increase the values of $T$ and $b_{defens}$.

The calibrated model parameters and the resulting $U$ values are given in Tables 4 and the first column of Table 5, respectively. Due to the stochastic nature of the model, separate runs of simulation with the optimal model parameters lead to different $U$ values. Nevertheless, we found that repeated runs only lead to slightly different $U$ values. All the simulated speed series show a good agreement with the empirical data, see Fig.12(c), which is the result of one run. Table 6 is the average time headway of the I-80 trajectory data of NGSIM collected at the location between station 7 and 8 on April 13, 2005. The average time headway varies between 1.78 *s* to 13.13 *s* in Tab.6. Since the average time headway of our calibration result is 4.5*s*, the value of $T$ is reasonable.

**Table 4**

Model parameters of NH model calibrated to the I-80-North detector data of 07. April 2005 by minimizing Theil's U.

| Parameters | $L_{cell}$ | $L_{veh}$ | $v_{max}$ | $T$ | $b_{defens}$ | $p_a$ | $p_b$ | $p_c$ | $g_{safety}$ | $t_c$ |
|---|---|---|---|---|---|---|---|---|---|---|
| Units | m | $L_{cell}$ | $L_{cell}/s$ | s | $L_{cell}/s^2$ | - | - | - | $L_{cell}$ | s |
| Value | 1 | 7 | 21 | 5.2 | 2 | 0.95 | 0.55 | 0.1 | 4 | 8 |

**Table 5**

Calibration (07. April 2005) and validation errors (other days) corresponding to Fig.12-17.

| Day | 07 Apr | 08 Apr | 11 Apr | 12 Apr | 13 Apr | 14Apr |
|---|---|---|---|---|---|---|
| $U$ | 0.0647 | 0.0689 | 0.0503 | 0.0485 | 0.0471 | 0.0653 |
| $U^M$ | 0.0362 | 0.1184 | 0.0182 | 0.0003 | 0.0006 | 0.0167 |
| $U^S$ | 0.0125 | 0.0218 | 0.0218 | 0.0036 | 0.0390 | 0.0343 |

**Table 6**

Average time headways (unit: s) by time period and lane (in seconds) of I-80 trajectories data.

| Time period (minutes) | lane | | | | |
|---|---|---|---|---|---|
| | 1 | 2 | 3 | 4 | 5 |
| 3:58:55-4:00 | 4.92 | 4 | 7.06 | 3.66 | 4.1 |
| 4:00-4:05 | 2.4 | 2.69 | 2.7 | 2.89 | 2.95 |
| 4:05-4:10 | 2.48 | 3.38 | 4.84 | 3.83 | 3.64 |
| 4:10-4:15 | 2.36 | 2.99 | 3.49 | 4.7 | 3.81 |
| 4:15-4:15:37 | 2.06 | 2.84 | 3.07 | 3.75 | 4.61 |
| 4:59:27-5:00 | 2.54 | 2.34 | 2.34 | 2.53 | 2.76 |
| 5:00-5:05 | 2.39 | 3.05 | 3.82 | 5.51 | 3.09 |
| 5:05-5:10 | 2.13 | 3.73 | 3.93 | 4.15 | 4.03 |
| 5:10-5:15 | 2.3 | 7.41 | 6.45 | 10.09 | 8.04 |
| 5:15-5:15:47 | 1.87 | 3.96 | 3.73 | 2.88 | 3.32 |
| 5:12:45-5:15 | 1.78 | 3.16 | 4.01 | 6.05 | 5.16 |
| 5:15-5:20 | 2.35 | 5.74 | 4.57 | 3.97 | 3.14 |
| 5:20-5:25 | 2.31 | 6.67 | 5.18 | 6.09 | 6.75 |
| 5:25-5:30 | 2.3 | 7.37 | 8.33 | 7.34 | 7.02 |
| 5:30-5:32:14 | 2.2 | 7.5 | 8.66 | 8.58 | 13.13 |

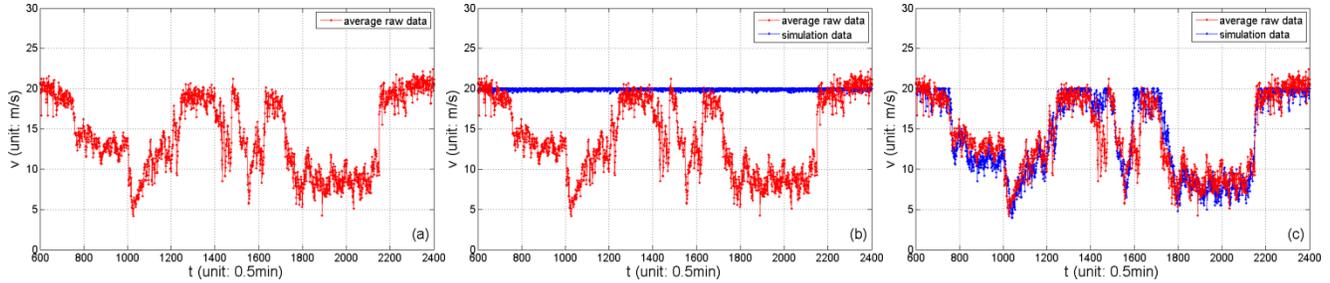

**Fig.12.** Time series of speed at Station 5 on Thu, 07 Apr 2005. (a) is the lane average speed series calculated by Equation (5). (b, c) are the comparisons between the data in (a) with the simulation speed series. (b) is obtained by the model with the parameter set table 3 with the revised cell length $L_{cell}= 1m$, maximum speed $v_{max}= 20L_{cell}/s = 20m/s$ and vehicle length $L_{veh}= 7L_{cell}= 7m$. (c) is obtained by the model with the parameter set Table 4.

*5.4. Validations*

In order to study the robustness of the calibrated parameters, we cross-validate the model by the data collected on other different days. All results are described in Tab.5 and the result of Fri, 08 Apr 2005 is shown in Fig.13. The model can capture the empirical traffic dynamics accurately on all days (08-12 Apr). All validation results are acceptable and better than that of the models tested by Brockfeld et al. (2005) and Wagner et al. (2010). They have tested many microscopic and macroscopic models on the same location with the same detectors at the opposed direction. The models include the NaSch model (Nagel and Schreckenberg, 1992), Newell's model (Newell, 2002), the OV model (Bando et al., 1995), the Cell Transmission model (Daganzo, 1994), Gipps's model (Gipps, 1981), the SK model (Krauss et al., 1997), the IDM (Treiber et al., 2000), and the macroscopic model proposed by Aw and Rascle (2000). The best model tested by Wagner et al. (2010) is the SK model, which was also tested by Brockfeld et al. (2005). The calibration and validation errors (U values) presented by Brockfeld et al. (2005) are in the range of 0.14 to 0.16, 0.14 to 0.23, respectively.

Moreover, it should be noted that our simulations are based on the homogeneous traffic and the heterogeneity is not considered, while real traffic flow is heterogeneous. Thus, the validation results mean that the real heterogeneous traffic can be simulated by the homogeneous traffic of NH model, which is highlighted as the one of the prominent advantages of the models within the three-phase theory (Kerner, 2012).

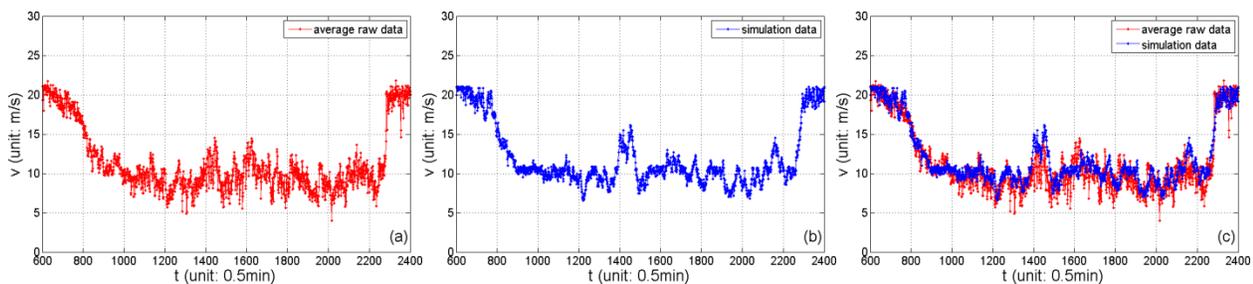

**Fig.13.** Time series of speed at Station 5 on Fri, 08 Apr 2005.(a) is the lane average speed series calculated by Equation (5). (b) is the simulation speed series. (c) is the comparisons between the data in (a) with the data in (b).

## 6. Conclusion

The fundamental diagram approach assumes the existence of a unique space gap vs. speed relationship, while the three-phase theory presumes that, within a certain range, drivers can make arbitrary choices of the space gap. In order to determine whether the unique space-gap-speed relationship exists, the US-101 trajectory datasets of NGSIM are analyzed. Results showed the following findings in 82% of the cases: (1) a linear relationship between actual space gap and speed can be identified when the speed difference between vehicles is approximately zero; (2) vehicles accelerate or decelerate around the desired space gap most of the time. To explain these phenomena, an assumption and a new cellular automaton model (NH model) are proposed such that, for homogeneous congested traffic flow in the noiseless limit, the space gap will oscillate around the desired space gap rather than keeping it exactly. This provides a possible dynamical explanation for the observed variation of the gaps for a given speed.

Two parts of simulations are conducted. In the first part, simulations on both a circular road and an open road with an on-ramp were carried out for NH model. Results obtained under the periodic conditions show that the NH model could produce the synchronized flow and two kinds of phase transitions which can be identified as F→S and S→J transitions. Results obtained from an open road with an on-ramp show that multiple congested patterns observed by simulating models of three-phase theory can be well reproduced by the NH model. In the second part, the NH model has been calibrated and validated by the I-80 detector datasets of NGSIM. Results show that the empirical data can be well reproduced and the validation errors are smaller than that of previous studies.

## Acknowledgements:

The authors wish to thank NGSIM for supplying the empirical data used in this article. Tian sincerely thanks for the help of Guojun Jiang and Yang Xu to deal with the empirical data. This work is supported by the National Natural Science Foundation of China (Grant Nos. 71271150, 71101102, 71222101, 71131001), and the 973 Program (No. 2012CB725400).

## Appendix: The three-phase theory

*A.1. Congested traffic phases*

In three-phase theory, congested traffic has been divided into the synchronized flow and wide moving jam phases, which are defined through empirical criteria [S] and [J]:

Wide moving jams [J]: A wide moving jam is a moving jam that maintains the mean speed of the downstream jam front, even when the jam propagates through other traffic phases or bottlenecks. Within the downstream front of the wide moving jam, vehicles accelerate from the standstill inside the jam to free flow. Within the wide moving jam, the vehicles are almost in a standstill or if they are moving, their speeds are very low. Within the upstream front of the wide moving jam vehicles must slow down to the speed within the jam. Generally, if the width of a moving jam (in the longitudinal direction) considerably exceeds the width of the jam fronts, one could call it a wide moving jam.

The synchronized flow phase [S]: In contrast with the wide moving jam phase, the downstream front of the synchronized flow phase does not exhibit the wide moving jam characteristic feature; in particular, the downstream front of the synchronized flow phase is often fixed at a bottleneck. In synchronized flow, the average speed of vehicles is noticeably lower and the density of vehicles is noticeably higher than the corresponding values in free traffic at the same flux of vehicles.

The criterion [J] could be explained by a flow interruption effect within a wide moving jam that occurs when vehicles are in a standstill or move with negligible low speed within the jam. A sufficient criterion for this flow interruption effect is:

$$T_{max} \gg T_{del}^{(ac)} \tag{9}$$

where $T_{max}$ is the maximum time headway between two vehicles within the jam and $T_{del}^{(ac)}$ is the mean time delay in vehicle acceleration at the downstream jam front from a standstill state within the jam. In a hypothetical case, when all vehicles within a moving jam do not move, the criterion for this flow interruption effect is:

$$T_J \gg T_{del}^{(ac)} \tag{10}$$

where $T_J$ is the jam duration, i.e. the time interval between the upstream and downstream jam fronts passing a detector location.

Condition (9) indicates that vehicles inside the moving jam is at least once in a stop during a large time interval compared with the mean time delay in vehicles acceleration from standstill at the downstream front. Under condition (9), there are at least several vehicles within the jam that are in a standstill or if they are still moving, it is only with a negligible low speed in comparison with the speed in the jam inflow and outflow. These vehicles could separate vehicles accelerating at the downstream jam front from vehicles decelerating at the upstream jam front. Therefore the jam inflow has no influence on the jam outflow, and the jam outflow only depends on the vehicles that accelerating from standstill at the downstream front. Thus, the traffic flow interruption effect can beused as a criterion to distinguish the synchronized flow from wide moving jams in single vehicle data.

*A.2. The fundamental hypothesis of the three-phase traffic theory*

The fundamental hypothesis of the three-phase theory is as follows: the hypothetical steady states of the synchronized flow cover a two-dimensional region in the flow-density plane, i.e., there is no fundamental diagram of traffic flow in this theory, Fig. A.1. The steady state of synchronized flow is a hypothetical state of synchronized flow of identical vehicles and drivers in which all vehicles move with the same time independent speed and have the same space gaps, i.e., this synchronized flow is homogeneous in time and space. This fundamental hypothesis assumes that the driver can make an arbitrary choice in the space gap to the preceding vehicle within a finite range of space gaps at a given speed in the steady states of synchronized flow.

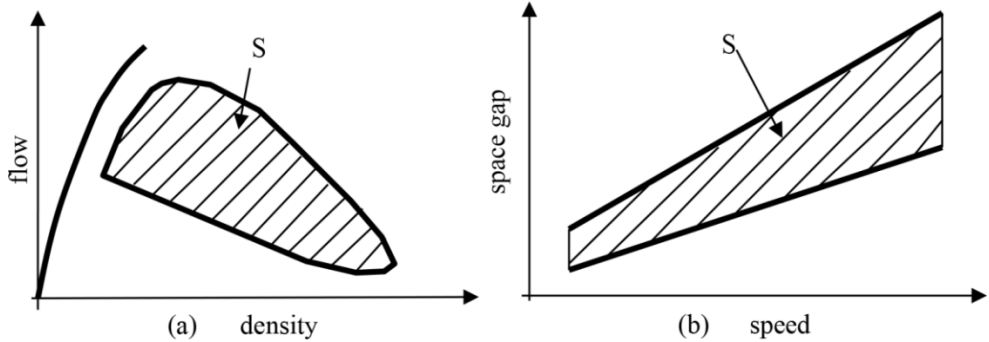

**Fig. A.1.** Fundamental hypothesis of three-phase traffic theory: (a) Qualitative representation of free flow states (F) and 2D steady states of synchronized flow (dashed region S) on a multi-lane road in the flow-density plane. (b) A part of the 2D steady states of synchronized flow shown in (a) in the space-gap-speed plane (dashed region S) (Kerner, 2009).

*A.3. Phase transitions*

In three-phase traffic theory, traffic breakdown is a phase transition from free flow to synchronized flow (F→S transition). Wide moving jams can occur spontaneously in synchronized flow only (S→J transition), i.e. due to a sequence F→S→J transitions. In real traffic, a fluctuation, whose amplitude exceeds the critical amplitude, occurring in the vicinity of the bottlenecks in the free traffic flow, will lead to the transition from free flow to synchronized flow (F→S transition). Jams emerge in the synchronized flow, i.e. narrow moving jams, who spontaneously emerge in the synchronized flow, move and grow in the upstream direction. Finally, these narrow moving jams (or a part of them) transform into wide moving jams (S→J transition).

F→S transition: if before traffic breakdown occurs at a bottleneck, there is free flow at the bottleneck as well as upstream and downstream in a neighborhood of the bottleneck, then the F→S transition is called as spontaneous F→S transition. On the other hand, if the F→S transition is induced by the propagation of a spatiotemporal congested traffic pattern, then the F→S transition is called as the induced F→S transition.

S→J transition: In empirical observations, S→J transition developments is associated with a pinch effect, which is the spontaneous emergence of growing narrow moving jams in the synchronized flow occurring within the associated pinch region of synchronized flow. If the growth of a nucleus required for moving jam emergence appears within the synchronized flow, the S→J transition is called as the spontaneous S→J transition. Just like the F→S transition, there also exists the induced S→J transition.

*A.4. Patterns at bottlenecks*

Empirical observations show that there are two main types of congested patterns at an isolated bottleneck:

The General Patterns (GPs): After the synchronized flow occurs upstream of the bottleneck, the wide moving jams continuously emerge in that synchronized flow and propagate upstream, and then this congested pattern is often called as the General Patterns (GP). However, if the wide moving jams discontinuous emerge on the road, there will be just one or few wide moving jams appearing in that synchronized flow, then this congested pattern is often called as the dissolving General Patterns (DGP).

The Synchronized Patterns (SPs): If there only is synchronized flow upstream of the bottleneck, no wide moving jams emerge in the synchronized flow, and then this congested pattern is often called as the Synchronized Patterns. And as a result of the F→S transition, various synchronized flow patterns can occurs at the bottleneck, such as the widening synchronized pattern (WSP), local synchronized pattern (LSP), moving synchronized pattern (MSP), and alternating synchronized pattern (ASP).

*A.5. Models based on the three-phase traffic theory*

In 2002, Kernerand Klenov (2002) proposed the KK car following model, which is able to show all known microscopic and macroscopic features of traffic breakdown, synchronized flow and congested patterns for the first time. Later, the one-lane KKW CA modeland two-lane KKS CA model (Kerner et al., 2002, 2011) are proposed. The main idea of above models is the speed adaptation effect within the synchronized distance. The vehicle tends to adjust its speed to the preceding vehicle as long as it is safe. Lee et al. (2004) developed the CA model mainly considering mechanical restriction versus human overreaction. This model could exhibit some features of SPs and GPs. The Brake Light CA Model (BLM, Knospe et al., 2000) and its variants (Comfortable Driving Models (CDMs)) (Jiang and Wu, 2003, 2005; Tian et al., 2009) have considered the brakelight effect, i.e., the simulated drivers adopt a more defensive driving strategy if the brake lights of the preceding vehicle are on, i.e., if this vehicle decelerates. The CDMsare based on theBLM. Simulation results of CDMsshow SPs and GPs as well as the diagram of congested patterns at an on-ramp bottleneck postulated in the three-phase traffic theory. The CA model by Gao et al. (2007, 2009) mainly assumes that randomization depends on speed difference. It is pointed out that this model is equivalent to a combination of the KKW model and the NaSch model. The car following model proposed by Davis (2004) incorporates the reaction delay into the optimal car following model, which can describe the F→S transition. The car following model by Kernerand Klenov (2006) considered different time delays on driver acceleration associated with driver behavior in various local driving situations, which can show spatiotemporal congested patterns that are adequate with empirical results. He et al. (2010) proposed a deterministic car-following model based on a multi-branch fundamental diagram with each branch representing a particular category of driving style. Traffic breakdown and some observed spatio-temporal patterns at on-ramp vicinity are reproduced.

In order to emphasize the significance of the two-dimensional steady states of synchronized flow, Kernerand

Klenov (2006) proposed the Speed Adaption Models (SAMs) in the framework of fundamental diagram approach. The basic hypothesis of SAMs is the double Z-characteristic for the sequence of phase transitions from free flow to synchronized flow to wide moving jams (F→S→J transitions). Based on this hypothesis, SAMs can reproduce both the traffic breakdown and the emergence of wide moving jams in synchronized flow as found in empirical observations. However, SAMs are not able to reproduce the local synchronized patterns (LSPs) consistent with empirical results as well as some of empirical features of synchronized flow between wide moving jams within general patterns (GPs). Kerner et al. attribute these drawbacks of SAMs to the lacking of the two-dimensional steady states of synchronized flow.